\documentclass[pra, showpacs, twocolumn, a4paper, nofootinbib]{revtex4}
\usepackage{graphicx}
\usepackage{subfigure}
\usepackage{amssymb}
\usepackage{amsmath}
\usepackage{bm}
\usepackage{latexsym}

\begin {document}

\newcommand{\ket}[1]{| #1 \rangle}

\title{Breakdown of the rotating-wave approximation in the description of entanglement of spin anticorrelated states}
\author{Jun Jing$^{1,2}$\footnote{Email: jungen@shu.edu.cn},
Zhi-Guo L\"{u}$^3$, and Zbigniew Ficek{$^4$}} 
\affiliation{$^1$Department of Physics, Shanghai University, Shanghai 200444, China \\
$^2$The Shanghai Key Lab of Astrophysics, Shanghai 200234, China\\
$^3$Department of Physics, Shanghai Jiao Tong University, Shanghai 200240, China \\
$^4$Department of Physics, The University of Queensland, Brisbane 4072, Australia}

\date{\today}

\begin{abstract}
It is well established that an entanglement encoded in the Bell states of a two-qubit system with correlated spins exhibits completely different evolution properties than that encoded in states with the anti-correlated spins. A complete and abrupt loss of the entanglement, called the entanglement sudden death,  can be found to occur for the spin correlated states, but the entanglement evolves without any discontinuity or decays asymptotically for the spin anti-correlated states. We consider the evolution of an initial entanglement encoded in the spin anti-correlated states and demonstrate that the asymptotic behavior predicted before occurs only in the weak coupling limit or equivalently when the rotating-wave approximation (RWA) is made on the interaction Hamiltonian of the qubits with the field. If we do not restrict ourselves to the RWA, we find that the entanglement undergoes a discontinuity, the sudden death phenomenon. We illustrate this behavior by employing an efficient scheme for entanglement evolution between two cold-trapped atoms located inside a single-mode cavity.  Although only a single excitation is initially present in the system, we find that the two-photon excited state, which plays the key role for the discontinuity  in the behavior of the entanglement, gains a population over a short time of the evolution. When the RWA is made on the interaction, the two-photon excited state remains unpopulated for all times and the discontinuity is absent. We attribute this phenomenon to the principle of complementarity between the evolution time and energy, and the presence of the counter-rotating terms in the interaction Hamiltonian.
\end{abstract}


\maketitle


An understanding of entanglement evolution and entanglement transfer between qubits is of fundamental interest in quantum information processing~\cite{Nielsen,Vedral}. The controlled transfer that preserves initial entanglement is crucially important. Transfer processes are susceptible to decoherence and dissipation due to the inevitable coupling of the qubits and the transfer channels to an external environment. Therefore, in order to minimize decoherence effects and to achieve the perfect fidelity, fast transfer processes or transfer operations performed over a very short time scale are highly desirable.

It has been recognized that the entanglement evolution depends on the state in which it is encoded. For a simple system of two qubits, the basis states for entanglement are four mutually orthogonal Bell states~\cite{ft02}. The states can be divided into two groups, one involving linear superpositions of the spin correlated states and the other involving spin anti-correlated states. The states belonging to these groups are often called two-photon and one-photon entangled states, respectively. The qubits can be prepared in a  spin correlated state by a transfer of  two-photon entangled states from quantum-correlated light fields produced e.g., in a nonlinear process of parametric down conversion~\cite{fd2}. Preparation of a spin anti-correlated state is more sophisticated as it involves a single excitation "shared" between two qubits. In principle, it can be achieved, for example by applying a short single laser pulse either in a running or in a standing wave configuration~\cite{be}. This will result in the qubits prepared either in a symmetric or in an antisymmetric combination of the spin anti-correlated states. 

Dynamics of an entanglement encoded in spin correlated states have been extensively studied since the pioneering work of Yu and Eberly~\cite{Yu,ey07},  who showed that an initial entanglement encoded in two separate qubits interacting with local environments can decay to zero in a finite time~\cite{Jing2}.
When the qubits are subjected of the interaction with each other through the coupling to the same environment, the already dead entanglement may revival after a finite time~\cite{ft06}. The interaction between the qubits induces a population difference between the symmetric and antisymmetric combinations of the spin anti-correlated states, which results in a nonzero entanglement. 

A completely different conclusion applies to the evolution of entanglement initially encoded in a spin anti-correlated state. It was pointed out by Jamr\'oz~\cite{j06} that an initial entanglement encoded in a spin anti-correlared state of two independent qubits interacting with local environments decays asymptotically in time without any discontinuity. This prediction agrees with Yonac and Eberly's work~\cite{yy07}, and also with other results~\cite{bf07}. The same conclusion applies to the case of two qubits mutually interacting through the coupling to a common environment~\cite{ft08}. In these papers, the analysis were restricted to the RWA and this raises the question on the validity of the results in the strong coupling regime where the breakdown in the RWA occurs~\cite{Cohen}.

In contrast to the numerous publications concerning atomic dynamics under the RWA, there are only a limited number of studies beyond the RWA. In terms of applications, the non-RWA calculations exploit short time or high-intensity dynamics where interesting effects appear not observed when the RWA is introduced. In this connection, we should mention the work on the Bloch-Siegert shift~\cite{bs40}, chaos in the Jaynes-Cummings model~\cite{ma83}, bifurcations in the phase space~\cite{p91} and a fine structure in the optical Stern-Gerlach effect~\cite{l08}. In the connection to entanglement, correlations between two separate atomic ensembles have recently been analyzed by Ng and Burnett~\cite{nb08}. It has been demonstrated that the ensembles interacting with the cavity field and initially prepared in their ground states undergo time evolution and become entangled over a short time only if the RWA is not made on the interaction Hamiltonian.

In this paper, we consider dynamics of the spin anti-correlated states of two identical qubits without the presence of any external excitations. We treat the problem fully quantum mechanically and do not restrict ourselves to the RWA. We find significant quantitative deviations from the RWA for the time evolution of an initial entanglement encoded in the spin anti-correlated states. We show that the entanglement may undergo a discontinuity that, on the other hand, requires either an initial or a transient  buildup of the population in the two-photon state of the system. One could argue that this rules out the discontinuity in the entanglement evolution since only a single excitation was present initially. We shall demonstrate that this is not the case if one considers an evolution with the non-RWA Hamiltonian and interpret this result is a consequence of the principle of complementarity between the evolution time and energy.

To describe the entanglement evolution in a two qubit system, we use concurrence, an entanglement measure that relates entangled properties to the coherence properties of the qubits~\cite{woo}.
In order to compute the concurrence, one needs the density matrix of the two qubit system written in the basis of the product states $\ket 1 =\ket{\!\downarrow\downarrow}, \ket 2 =\ket{\!\uparrow\downarrow}, \ket 3 =\ket{\!\downarrow\uparrow}, \ket 4 =\ket{\!\uparrow\uparrow}$. 
Here, $\ket{\!\uparrow\downarrow}$ represents the qubit $1$ in the excited ({}``up'') state, the qubit $2$ in the ground ({}``down'') state.
The density matrix is in general composed of sixteen nonzero elements. We make an assumption that the qubits evolve without the presence of any external fields. In this case, the density matrix takes a simple block diagonal form
\begin{eqnarray}
  \rho(t) = \left(
    \begin{array}{cccc}
      \rho_{11}(t) & 0 & 0 & 0 \\
      0 & \rho_{22}(t) & \rho_{23}(t) & 0\\
      0 & \rho_{32}(t) & \rho_{33}(t) & 0\\
      0 & 0 & 0 &\rho_{44}(t)
    \end{array}\right) ,\label{e1}
\end{eqnarray}
in which we keep all the diagonal elements (populations) and the coherences $\rho_{23}(t)$ 
and $\rho_{32}(t)$ that might be nonzero initially or can build up during the evolution of the system. 

For a system described by the density matrix (\ref{e1}), the concurrence has the form
\begin{eqnarray}
  C(t) = \max\left\{0,\, {\cal C}(t)\right\} ,\label{e2} 
\end{eqnarray}
with
\begin{eqnarray}
  {\cal C}(t) = 2|\rho_{23}(t)| -2\,\sqrt{\rho_{11}(t)\rho_{44}(t)} ,\label{e3} 
\end{eqnarray}
from which we see immediately that the discontinuity behavior cannot be achieved in any state of the system if only a single excitation was present since $\rho_{44}(t)=0$. In this case, the concurrence depends solely on the coherence $\rho_{23}(t)$. The discontinuity or threshold behavior of the concurrence requires a nonzero population of the ground and the upper two-photon states of the system. If an initial excitation were redistributed during the evolution among all of the states, we would have a possibility for a discontinuous evolution of entanglement.  We then would have a prototype of the entanglement sudden death. 

To gain further insight into the discontinuity behavior of entanglement, we employ the Bell states that are maximally entangled states of two qubits with correlated and anti-correlated spins
\begin{eqnarray}
|s\rangle &=& \frac{1}{\sqrt{2}}\!\left(|\!\uparrow\downarrow\rangle 
+|\!\downarrow\uparrow\rangle\right) ,\
|a\rangle = \frac{1}{\sqrt{2}}\!\left(|\!\uparrow\downarrow\rangle 
-|\!\downarrow\uparrow\rangle\right) ,\nonumber \\
|\alpha\rangle &=& \frac{1}{\sqrt{2}}\!\left(|\!\uparrow\uparrow\rangle 
+|\!\downarrow\downarrow\rangle\right) ,\ 
|\beta\rangle = \frac{1}{\sqrt{2}}\!\left(|\!\uparrow\uparrow\rangle 
-|\!\downarrow\downarrow\rangle\right) ,\label{e4}
\end{eqnarray}
In terms of the Bell states, the concurrence (\ref{e3}) has the form
\begin{eqnarray}
  {\cal C}(t) &=& \sqrt{\left[\rho_{ss}(t)\!-\!\rho_{aa}(t)\right]^{2}\!-\!\left[\rho_{sa}(t)\!-\!\rho_{as}(t)\right]^{2}} \nonumber \\
 &-& \sqrt{\left[\rho_{\alpha\alpha}(t)\!+\!\rho_{\beta\beta}(t)\right]^{2}\!-\!\left[\rho_{\alpha\beta}(t)\!+\!\rho_{\beta\alpha}(t)\right]^{2}} .\label{e5} 
\end{eqnarray}
As the entanglement has been defined by the requirement that ${\cal C}(t)>0$, it follows immediately from Eq.~(\ref{e5}) that the entanglement depends on the distribution of the population between the spin correlated and anti-correlated states. An entanglement creation involving the spin anti-correlated states is diminished by the presence of population and coherence between the spin correlated states. The evident competition between the spin correlated and anti-correlated states in the creation of entanglement may lead to a discontinuity in the time evolution of the entanglement. Therefore it immediately raises the question of whether a discontinuity in the evolution of the entanglement can ever be achieved if initially only a single excitation was present and no external excitations are applied to the system. This problem is treated quantitatively below. 

Let us now examine the above general results for the properties of the concurrence on a simple example of two identical two-level atoms (qubits) located inside a high-$Q$ single-mode standing-wave cavity~\cite{rb01}. We assume that the qubits undergo a strong coupling to the cavity mode when located at the antinode of the cavity field, and consider the situation where the strength of the coupling is controlled by varying the position of the qubits inside the standing-wave cavity mode.
The dynamics of the system are determined by the master equation of the density operator $\rho_{s}$ of the total, qubits plus the cavity field system
\begin{equation}
\frac{\partial\rho_{s}}{\partial t}= -\frac{i}{\hbar}[H,\rho_{s}] 
- \frac{1}{2}\kappa\left(a^\dagger a\rho_{s} + \rho_{s} a^\dagger a -2a\rho_{s} a^\dagger\right) ,\label{e6}
\end{equation}
where $\kappa$ is the damping rate of the cavity mode, and
\begin{eqnarray}
H_{{\rm nRWA}} &=& \frac{1}{2}\hbar\omega_{0}\sum_{j=1}^2\sigma^z_j+\hbar\omega a^\dagger a \nonumber \\
&& +\hbar\sum_{j=1}^2\left[g(r_{j})\sigma^{x}_{j}a^\dagger +g^{\ast}(r_{j})a\sigma^{x}_{j}\right] \label{e7}
\end{eqnarray}
is the non-RWA Hamiltonian of the qubits of transition frequency $\omega_{0}$ interacting with the single-mode cavity field of frequency $\omega$. The operators $a$ and $a^{\dagger}$ are the annihilation and creation operators of the cavity field, while $\sigma_{x}$ and $\sigma_{z}$ are the Pauli matrices in the $x$ and $z$ directions, respectively.
The parameter $g(r_{j})$ is the coupling constant between the $j$th qubit and the cavity mode at a position $r_{j}$ of the qubit along the cavity axis. The coupling constant varies with the position of the qubits~as
\begin{eqnarray}
g(r_{1,2}) = g_{0}\sin[\pi(L\mp d)/\lambda] ,\label{e8} 
\end{eqnarray}
where $L$ is the size of the cavity and $\lambda$ is the cavity wavelength. We assume that the qubits are placed symmetrically about the antinode of the cavity mode such that $r_{1}+r_{2}=L$ and $r_{2}-r_{1}=d$ is the distance between the qubits. When $d=0$, the qubits are then at the antinode of the cavity mode and experience the peak coupling strength $g_{0}$. As $d$ increases, the coupling strength decreases and approaches zero at $d=L$.


When the RWA is made, the Hamiltonian (\ref{e7}) takes the form
\begin{eqnarray}
H_{{\rm RWA}} &=& \frac{1}{2}\hbar \omega_{0}\sum_{j=1}^2\sigma^z_j
+\hbar\omega a^\dagger a \nonumber \\
&& +\sum_{j=1}^2\left[g(r_{j})\sigma^-_ja^\dagger +g^{\ast}(r_{j})a \sigma^\dagger_j \right] ,\label{e9}
\end{eqnarray}
where the counter-rotating terms have been ignored. The RWA is valid if the cavity field is nearly resonant with the atomic transition frequency, $\omega\approx\omega_{0}$, and the coupling constant $g(r_{j})$ is much smaller than $\omega$, i.e., $g(r_{j})\ll \omega$.

The master equation (\ref{e6}) can be solved numerically for various values of $L$ and $d$ and for different initial conditions. For our purposes here, it is sufficient to focus on the initial condition of the two qubits prepared in the spin anti-correlated state $\ket s$. We calculate the concurrence between the atoms by tracing the density operator~$\rho_{s}$ over the cavity field assumed to be in the vacuum 
state~$\ket 0$. 
\begin{figure}[h]
\includegraphics[width=3in]{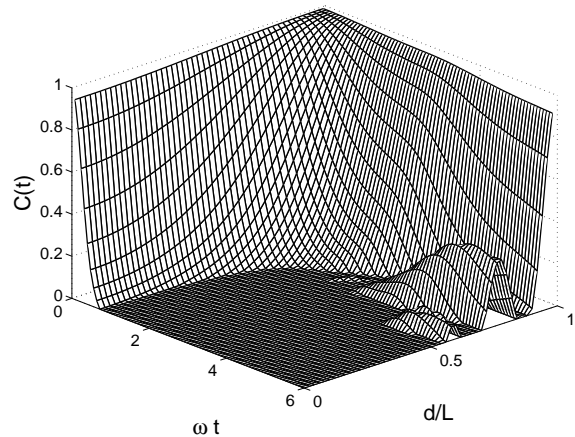}
\caption{Concurrence as a function of the scaled time $\omega t$ and distance between the qubits evaluated without the RWA for  $g_{0}=\omega$, $(\omega -\omega_{0})/\omega =0.01$, $\kappa =0.1\omega$ and $L=\lambda/2$.} 
\label{fig2}
\end{figure}

Consider first the evolution of an initial entanglement without the RWA.
Figure~\ref{fig2} displays the time evolution of the concurrence for the initial spin anti-symmetric state $\ket s$ and for the case of a strong coupling of the qubits to the cavity mode, $g_{0}=\omega$. We see that the initial entanglement undergoes the discontinuity, the sudden death behavior. The entanglement disappears quite rapidly over a very short evolution time, $\omega t\approx 1$. The sudden death behavior appears for small distances between the qubits, where a strong coupling is present, $g(r_{j})$ not much different from $g_{0}$. For large distances, where $g(r_{j})$ are very small, the discontinuity disappears and the entanglement decays asymptotically in time, the behavior predicted before under the RWA. 

The entanglement sudden death seen in Fig.~\ref{fig2} seem puzzling at first, because it appears to contradict the predictions based upon a simple argument that with a single excitation present in the system, it is impossible to achieve the discontinuity as, according to Eq.~(\ref{e3}), it requires population of the two-photon state. To resolve this problem, we now discuss the origin of the population of 
the two-photon state. We offer two complimentary views of the underlying physics. The first is provided by the derivation of the reduced density operator for the atoms using state vectors of the combined atoms plus the cavity field system. Assume that there is only a single excitation in the system that evolves under the~RWA Hamiltonian. In this case, the space of the system is spanned by three-state vectors, $|\!\uparrow\downarrow\rangle |0\rangle$, $|\!\downarrow\uparrow\rangle |0\rangle$ and 
$|\!\downarrow\downarrow\rangle |1\rangle$, where $|0\rangle$ and $|1\rangle$ are the zero-photon and one-photon Fock states the cavity mode.

If we evaluate the reduced density operator $\tilde\rho$ for the atoms by tracing over the cavity 
mode, we find
\begin{eqnarray}
\tilde{\rho} &=& {\rm Tr}_{cavity}\rho=\sum_{i=0,1}\langle i|\rho|i\rangle 
= \!\rho_{11}(t)|\!\downarrow\downarrow\rangle\langle \downarrow\downarrow\!|\nonumber \\
&+& \!\rho_{22}(t)|\!\uparrow\downarrow\rangle\langle \downarrow\uparrow\!|
+\!\rho_{33}(t)|\!\downarrow\uparrow\rangle\langle \uparrow\downarrow\!| \nonumber \\
&+& \!\rho_{23}(t)|\!\uparrow\downarrow\rangle\langle \uparrow\downarrow\!|
+\!\rho_{32}(t)|\!\downarrow\uparrow\rangle\langle \downarrow\uparrow\!| ,\label{e10}
\end{eqnarray}
where $|i\rangle$ refers to the cavity mode and $\rho_{ij}$ are populations $(i=j)$ of the atomic states and coherences $(i\neq j)$ between them.
Thus, under the RWA, the evolution of the atoms does not involve the two-photon (upper) atomic state.

Consider now the evolution of the system under the non-RWA Hamiltonian, that includes the counter-rotating terms $a \sigma^-_j$ and $\sigma^\dagger_j a^\dagger$. In this case, 
we must include the processes that do not strictly conserve excitation number, when one of the atoms goes to the excited (ground) state by emitting (absorbing) a photon. Thus, the space of the system is now spanned by six-state vectors, 
$|\!\uparrow\downarrow\rangle |0\rangle$, $|\!\downarrow\uparrow\rangle |0\rangle$,  
$|\!\downarrow\downarrow\rangle |1\rangle$, $|\!\uparrow\downarrow\rangle |2\rangle$, 
$|\!\downarrow\uparrow\rangle |2\rangle$ and $|\!\uparrow\uparrow\rangle |1\rangle$. 
This leads to the reduced density operator of the form
\begin{eqnarray}
\tilde{\rho} &=& \!\rho_{11}(t)|\!\downarrow\downarrow\rangle\langle \downarrow\downarrow\!|
+ \!\rho_{22}(t)|\!\uparrow\downarrow\rangle\langle \downarrow\uparrow\!| \nonumber \\
&+& \!\rho_{33}(t)|\!\downarrow\uparrow\rangle\langle \uparrow\downarrow\!| 
+ \!\rho_{44}(t)|\!\uparrow\uparrow\rangle\langle \uparrow\uparrow\!| \nonumber \\
&+&\!\rho_{32}(t)|\!\downarrow\uparrow\rangle\langle \downarrow\uparrow\!| 
+ \!\rho_{23}(t)|\!\uparrow\downarrow\rangle\langle \uparrow\downarrow\!| \nonumber \\
&+& \!\rho_{14}(t)|\!\downarrow\downarrow\rangle\langle \uparrow\uparrow\!| 
+ \!\rho_{41}(t)|\!\uparrow\uparrow\rangle\langle \downarrow\downarrow\!| .\label{e11}
\end{eqnarray}
It is evident from Eq.~(\ref{e11}) that the evolution of the atoms under the non-RWA Hamiltonian involves the two-photon state, that population of the state $|\!\uparrow\uparrow\rangle$ becomes possible.

A second explanation follows the principle of complementarity between the evolution time $\Delta t$ 
and uncertainty $\Delta E$ in the energy
\begin{eqnarray}
\Delta t\Delta E \geq \hbar .\label{e12}
\end{eqnarray}
For the evolution time of the order $1/\omega$, that is considered here, the uncertainty in the energy is of the order  $\hbar \omega$, the order required to achieve a non-zero population of the upper state 
$|\!\uparrow\uparrow\rangle$ of the two-atom system.
\begin{figure}[h]
\includegraphics[width=3in]{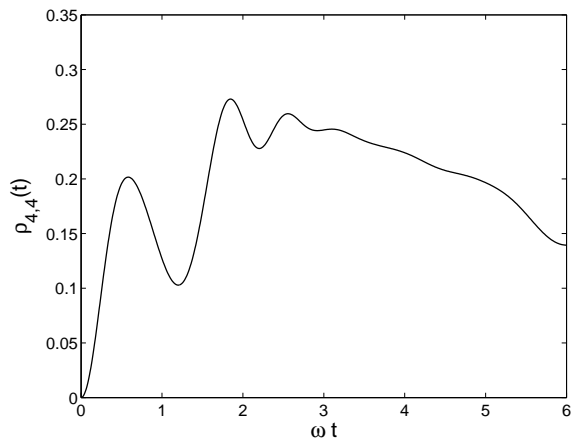}
\caption{Time evolution of the population of the two-photon state $\ket 4$ calculated without the RWA for $g_{0}=\omega$, $(\omega -\omega_{0})/\omega =0.01$, $\kappa =0.1\omega$, $L=\lambda/2$ and $d=0$.}
\label{fig3}
\end{figure}

To emphasize the participation of the two-photon state in the atomic dynamics under the non-RWA Hamiltonian,  we plot in Fig.~\ref{fig3} the time evolution of the population of the two-photon state 
$\ket 4 =\ket{\!\uparrow\uparrow}$ for the same parameters as in Fig.~\ref{fig2} with $d=0$. It is seen that initially unpopulated state $\ket 4$ becomes populated in a comparatively short time.

\begin{figure}[h]
\includegraphics[width=3in]{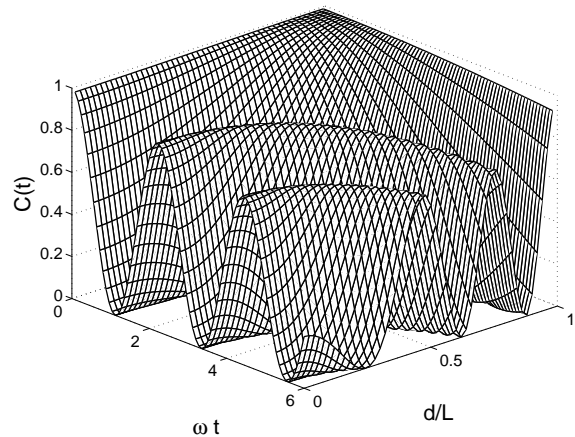}
\caption{Concurrence as a function of the scaled time $\omega t$ and distance between the qubits calculated with the RWA for $g_{0}=\omega$, $(\omega -\omega_{0})/\omega =0.01$, $\kappa =0.1\omega$ and $L=\lambda/2$.} 
\label{fig4}
\end{figure}

As we have already pointed out,  the two-photon state can be populated only if the counter-rotating terms are present in the interaction Hamiltonian. To show this expicitly, we look at the evolution of the entanglement under the RWA approximation. In Fig.~\ref{fig4}, we illustrate the time evolution of the concurrence for the same parameters as in Fig.~\ref{fig2}, but under the RWA Hamiltonian~(\ref{e9}). We see that in contrast to the non-RWA case, the entanglement evolves continuously without any discontinuity. Clearly, the counter-rotating terms are responsible for the entanglement sudden death. Under the strong coupling situation and small distances between the qubits, the entanglement oscillates with the Rabi frequency $g_{0}$. The oscillations disappear and the evolution ceased for large distances, where the coupling constants are small. The fact that the RWA version of the interaction precludes the entanglement sudden death suggests that caution should be exercised in the studies of entanglement evolution in a strong coupling limit. We stress that the coupling strengths considered in this papers have not been realized yet. However, there are proposals of realistic systems involving high $Q$ cavities~\cite{rb01,Im99,mm06} or nanomechanical resonators~\cite{sc04}, where such strong couplings could be realized experimentally with the current technology. 

In summary, we have considered the time evolution of an entanglement initially encoded in a spin anti-correlated state of two identical qubits. What is special with this entanglement, that no discontinuity behavior has been predicted under the RWA. We have demonstrated the failure of the RWA in the description of the evolution of entanglement in the strong coupling regime of the qubits to the field they interact with. The results show that the better the RWA is in describing the dynamics, the weaker the entanglement sudden death behavior, and only in the limit of a weak coupling, the RWA is an excellent approximation for the entanglement evolution of the spin anti-correlated states.

We would like to acknowledge the support from the National Natural Science Foundation of China under grant No. 10804069, Shanghai Education Foundation for Young teachers, the Shanghai Research Foundation No. 07dz22020 and the Australian Research Council.

\end{document}